\documentclass[preprint,showpacs]{revtex4-1}          
\usepackage{amsmath, amsfonts,natbib}
\usepackage{booktabs}
\usepackage{graphicx}
\usepackage{epsfig}
\usepackage{color, bm}
\usepackage{multirow}
\newcommand{\fig}[1]{Fig.~\ref{#1}}

\begin{document}
\title { Enhanced Photoabsorption from Cobalt Implanted Rutile TiO$_{2}$ (110) Surfaces}
\author{Shalik Ram Joshi$^1$, B. Padmanabhan$^2$, Anupama Chanda$^3$, Indrani Mishra$^1$, V. K. Malik$^2$, N. C. Mishra$^4$,
D. Kanjilal$^5$ and Shikha Varma$^1$}
\email{shikha@iopb.res.in}
\affiliation{$^1$Institute of Physics, Sachivalaya Marg, Bhubaneswar-751005, India\\ 
$^2$Department of Physics, Indian Institute of Technology, Roorkee 247667, India\\
$^3$Department of Physics, Dr. Hari Singh Gour University, M.P 470003, India\\
$^4$Department of Physics, Utkal University, Bhubaneswar-751004, India\\
$^5$Inter University Accelerator Center, New Delhi 110067, India}
\date{\today}
\begin{abstract}
Present study investigates the photoabsorption properties of single crystal rutile TiO$_{2}$ (110) surfaces after
they have been implanted with low fluence of Cobalt ions. The surfaces, after implantation,  demonstrate fabrication of
nanostructures and anisotropic nano-ripple patterns. Creation of oxygen vacancies (Ti$^{3+}$ states) as well
as band gap modification for these samples is also observed. Results presented here demonstrate that fabrication
of self organized nanostructures and development of oxygen vacancies, upon cobalt implantation, promote the enhancement
of photoabsorbance in both UV ($\sim$2 times) and visible ($\sim$5 times) regimes. These investigations on nanostructured
TiO$_{2}$ surfaces can be important for photo- catalysis.

\end{abstract}

\maketitle
\section{Introduction}
\label{introduction} 
Studies of metal oxide semiconductors like TiO$_{2}$, at nanoscales, have been of great interest for many years due to their 
numerous technological applications in photo- catalysis, solar cells, photovoltaics, magnetic storage media, waste water management,
etc. \cite{Fuj, Fuj1, Reg}. TiO$_{2}$ is a wide band gap semiconductor and under UV irradiation can produce hydroxyl radical 
(OH$^{-}$) which act as a powerful oxidizing agent to disintegrate  many organic pollutants dissolved in water \cite{Ino}. 
However, due to its large band gap (3.2 eV for rutile TiO$_{2}$) it absorbs visible light poorly, and thus has been nearly 
ineffective for visible light photo- catalysis. Various methods have been used to improve the photocatalytic activity
of rutile TiO$_{2}$, e.g. through dye sensitization, synthesis as thin films, formation of nanocrystals, incorporation of
dopants by chemical methods, etc. \cite{Par1, Tan, Wan}.  The enhancement of photocatalytic activity of rutile TiO$_{2}$
by most of the above methods is however somehow limited, as organic dyes can become unstable at high temperature and
the diffusion of dopants depend on the temperature and complexity of the chemical methods \cite{Cho, Don, Gut}.

Ion implantation is an important technique for introducing dopants in the host lattice. Implantation sometimes 
leads to the development of nanostructures on the surfaces. Such fabrication of self organized nano-patterns have
been observed on a variety of semiconductor and metal surfaces \cite{Val}. These patterns develop due to the competition
between the curvature dependent sputtering process, which erode the surface atoms, and various relaxation mechanisms 
\cite{Bra}. During implantation of bi-atomic surfaces, preferential sputtering of low mass atoms can also take place
which may lead to the development of vacancies as well as metal rich centers \cite{Sig}. 
 
The present study investigates the photoabsorption properties of rutile TiO$_{2}$(110) surfaces after their implantation 
with Co ions. The results show formation of nanostructured patterns as well as creation of oxygen vacancies, or Ti$^{3+}$ 
states, on the ion implanted  surfaces. For these ion implanted surfaces an enhanced photoabsorption in UV as well as
visible regimes is observed. The presented results demonstrate that development of self organized nanostructures, 
creation of vacancy states, and modification in band gap, upon Co ion implantation, leads to the observed increase in 
photoabsorption in TiO$_{2}$(110).

\section{Experimental Details}
\label{exp}
Commercially available rutile single crystal TiO$_{2}$ (110) substrates were irradiated with 200~keV cobalt ions, 
at room temperature, using 15~MV Pelletron Accelerator.  The Co ions, incident along the substrate normal, were
implanted in TiO$_{2}$ at a fluence of $3\times10^{16}~ions/cm^{2}$. The flux of the ion source was 
$1.6\times10^{13}~ions/cm^{2}~sec$. The range of Co atoms in TiO$_{2}$ has been estimated to be 98~nm by 
SRIM \cite{Zie}. Scanning Probe Microscope (SPM), Nanoscope V (Bruker), was utilized in 
tapping mode. For X-Ray Diffraction (XRD) measurements, a Bruker system equipped with Cu-K$_{\alpha}$ source ($\lambda\sim$ 0.154 nm) was utilized.
X-Ray Photoelectron Spectroscopy (XPS) studies were performed on a VG instrument which is equipped with a
Concentric Hemispherical Analyzer and a dual x-ray source (Mg-K$_{\alpha}$ and Al-K$_{\alpha}$). The 
resolution of the system is 0.9~eV. The spectra were acquired using an Mg-K$_{\alpha}$ source with a  
pass energy of 20~eV. Optical absorption studies were investigated using Shimadzu UV-vis spectrophotometer.

\section{Results and discussion}
\label{result}
\begin{figure}
\includegraphics[width=1.0\textwidth]{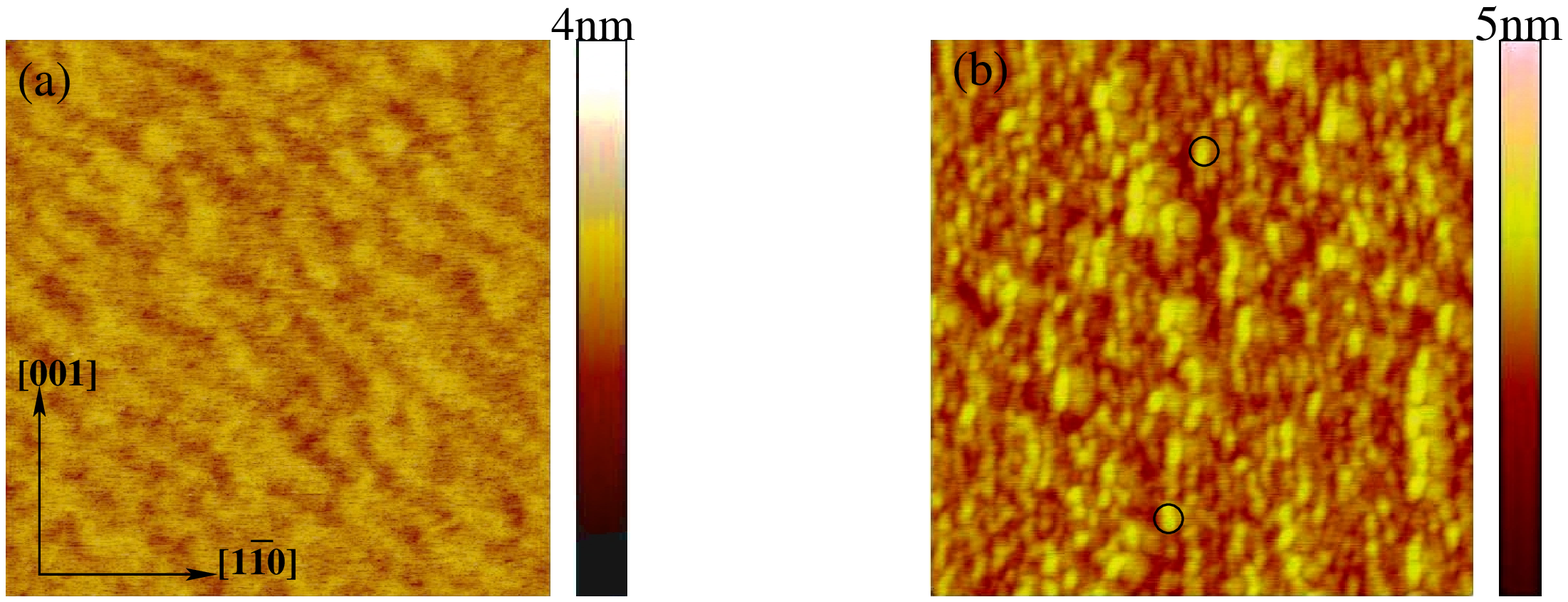}
\caption{AFM images $(1\mu m\times 1\mu m)$ of rutile TiO$_{2}$ for (a) pristine, and after (b) cobalt implantation. 
Crystallographic directions of the surface are shown on pristine surface and remain same for the
implanted sample. Few nanostructures, on the implanted surface, are marked with solid circles.}
\label{fig:afm}
\end{figure}

\begin{figure}
\includegraphics[width=1.0\textwidth]{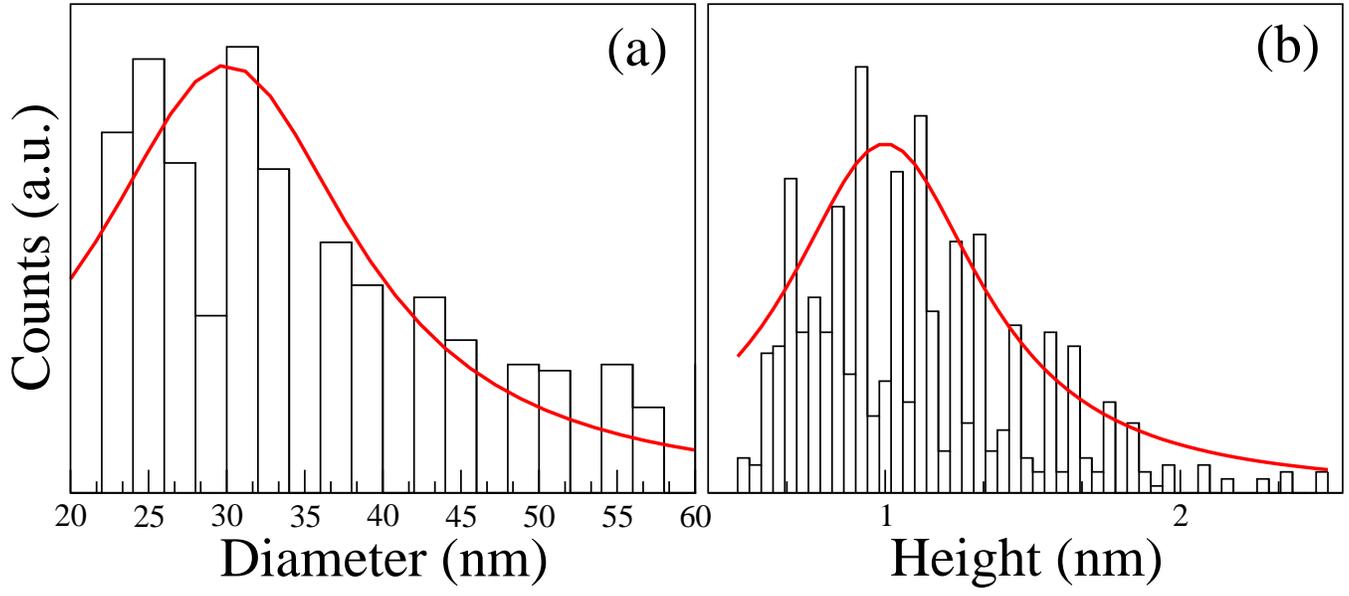}
\caption{(a) Diameter, and (b) Height distributions of nanostructures that develop on TiO$_{2}$ surface after cobalt implantation.  Solid 
lines represent Lorentzian fittings to the distributions.}
\label{fig:diameter}
\end{figure}

Morphological evolution of rutile TiO$_{2}$ surfaces is shown in \fig{fig:afm}. High resolution image from a pristine 
TiO$_{2}$ displays steps across the surface (\fig{fig:afm}(a)). The surface is smooth and shows a small mean
roughness of 0.006~nm. Similar stepped morphologies with smooth surfaces have been earlier observed for 
TiO$_{2}$ \cite{Ste}. The implanted surface, however, is observed to be decorated with a large density
($1.16\times10^{10}/cm^{2}$) of nanostructures (outline for some are marked) in \fig{fig:afm}(b). 
The diameter (d) and height (h) distributions of these nanostructures, with an average $<d>$ and $<h>$ of $\sim$30~nm 
and $\sim$1~nm respectively, are shown in \fig{fig:diameter}. These nanostructures predominantly display a 
2-dimensional nature, with ratio $\frac{<d>}{<h>}$ being nearly 30.

An important consequence of the ion impact on single crystal substrates is the creation of adatoms and vacancy clusters 
\cite{Val}. Along with the formation of well defined nanostructures, presence of ripple patterns 
can also be observed on the implanted TiO$_{2}$ surfaces in \fig{fig:afm}(b). 
These ripples run along the [001] crystallographic direction and their ripple wave- vector is along [1$\overline{1}$0] 
direction. The wavelength of these ripples is about 30~nm. The morphological evolution of the
surfaces, during ion irradiation, predominantly happens due to various competing processes. Erosion and curvature 
dependent sputtering etc. induce surface instability whereas the process of surface diffusion of adatoms maintains
the equilibrium \cite{Val}. 

\begin{figure}
\includegraphics[width=0.5\textwidth]{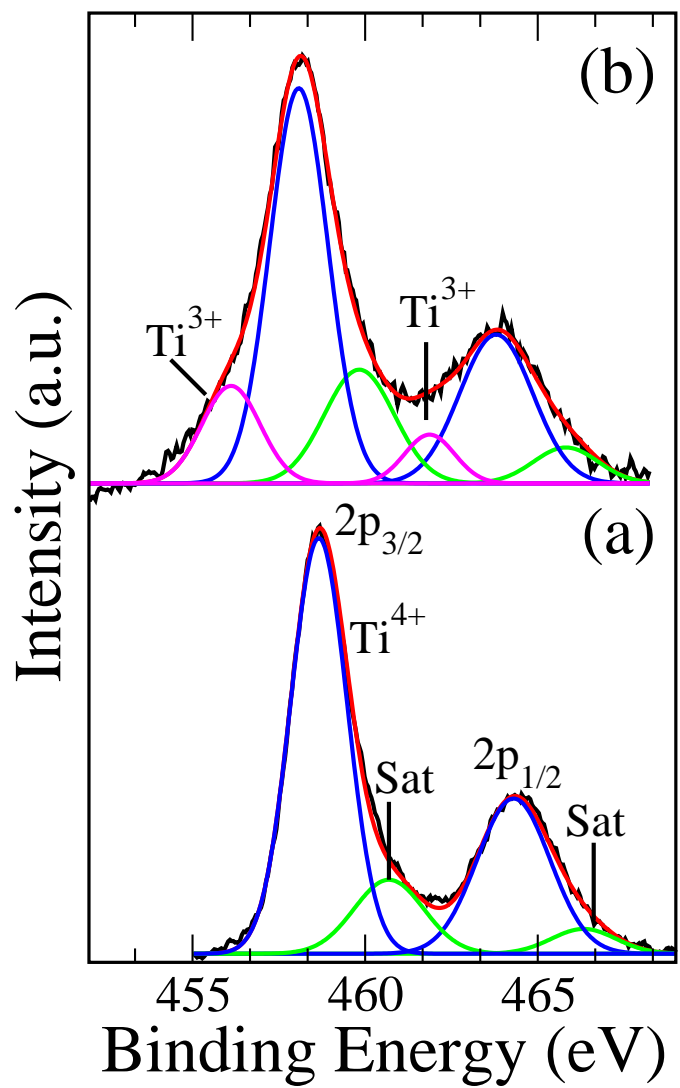}
\caption{XPS Core level spectra of Ti (2p) of rutile TiO$_{2}$ for (a) pristine surface, and after (b) cobalt implantation. Fittings of
components are shown.}
\label{fig:ti2p}
\end{figure}

XPS is a very surface sensitive technique, with the photoelectron signal primarily coming from top $\sim$1~nm layer of the 
TiO$_{2}$ \cite{Bri}. \fig{fig:ti2p} displays the Ti (2p) spectra for rutile TiO$_{2}$ from the pristine as well as Co implanted TiO$_{2}$.
XPS from pristine surface shows the presence of 2p$_{3/2}$ and 2p$_{1/2}$ features at 458.6~eV and 464.3~eV, respectively. These features
are related to the Ti$^{4+}$ coordinated sites on the rutile TiO$_{2}$ surface \cite{Sol, Maj}. Each of these states have an 
associated feature towards higher binding energy side (labeled as {\it Sat}). This is due to the shake up satellite from 
TiO$_{2}$ \cite{Kha, Mas}. No sign of impurity or any Ti$^{3+}$ vacancy state is observed for the pristine TiO$_{2}$.
\fig{fig:ti2p}(b) shows the XPS from the Co implanted TiO$_{2}$. Here, in addition to Ti$^{4+}$ related and 
shake- up- satellite related features, new features are also observed. These are due to the creation of oxygen vacancy states, 
or Ti$^{3+}$ states, which get created during the ion irradiation of TiO$_{2} $ \cite{Jeo}. These Ti$^{3+}$ related features for
2p$_{3/2}$ and 2p$_{1/2}$ states appear, respectively, at 456.1~eV and 461.8~eV in \fig{fig:ti2p}(b).

On hetroatomic crystals, preferential sputtering, during ion irradiation, usually occurs due to the difference in
the binding energies of various atoms. This can sometimes lead to the development of metal rich centers as well as 
formation of vacancy states. After Co implantation in TiO$_{2}$, the oxygen atoms get preferentially
sputtered as compared to titanium atoms and lead to the development of Ti$^{3+}$ states as seen in \fig{fig:ti2p}(b).
With sputtering of oxygen atoms, the associated electrons go to empty 3d orbital of neighboring Ti atom forming the two 
Ti$^{3+}$ sites.  This leads to the creation of Ti rich zones which become the nucleation centers for the development 
of nanostructures \cite{Sol, Maj}. Room temperature DFT studies on rutile TiO$_{2}$ (110) surface  have shown that the
diffusion barrier for vacancies as well as adatoms is lower along [001] crystallographic direction compared to 
the [1$\overline{1}$0] direction \cite{Kol}.  The formation of anisotropic ripples running along [001], as observed here 
in \fig{fig:afm}(b), can be related to the non- symmetric diffusion barriers for Ti$^{3+}$ vacancies. For metal 
substrates also similar results have been reported earlier \cite{Val, Sha}. The development of nanostructures as well
anisotropic ripple patterns, on ion implanted TiO$_{2}$ surfaces as observed in the present study, are crucially 
controlled by the preferential sputtering during irradiation and the non-symmetric diffusion of vacancies.

\begin{figure}
\includegraphics[width=1.0\textwidth]{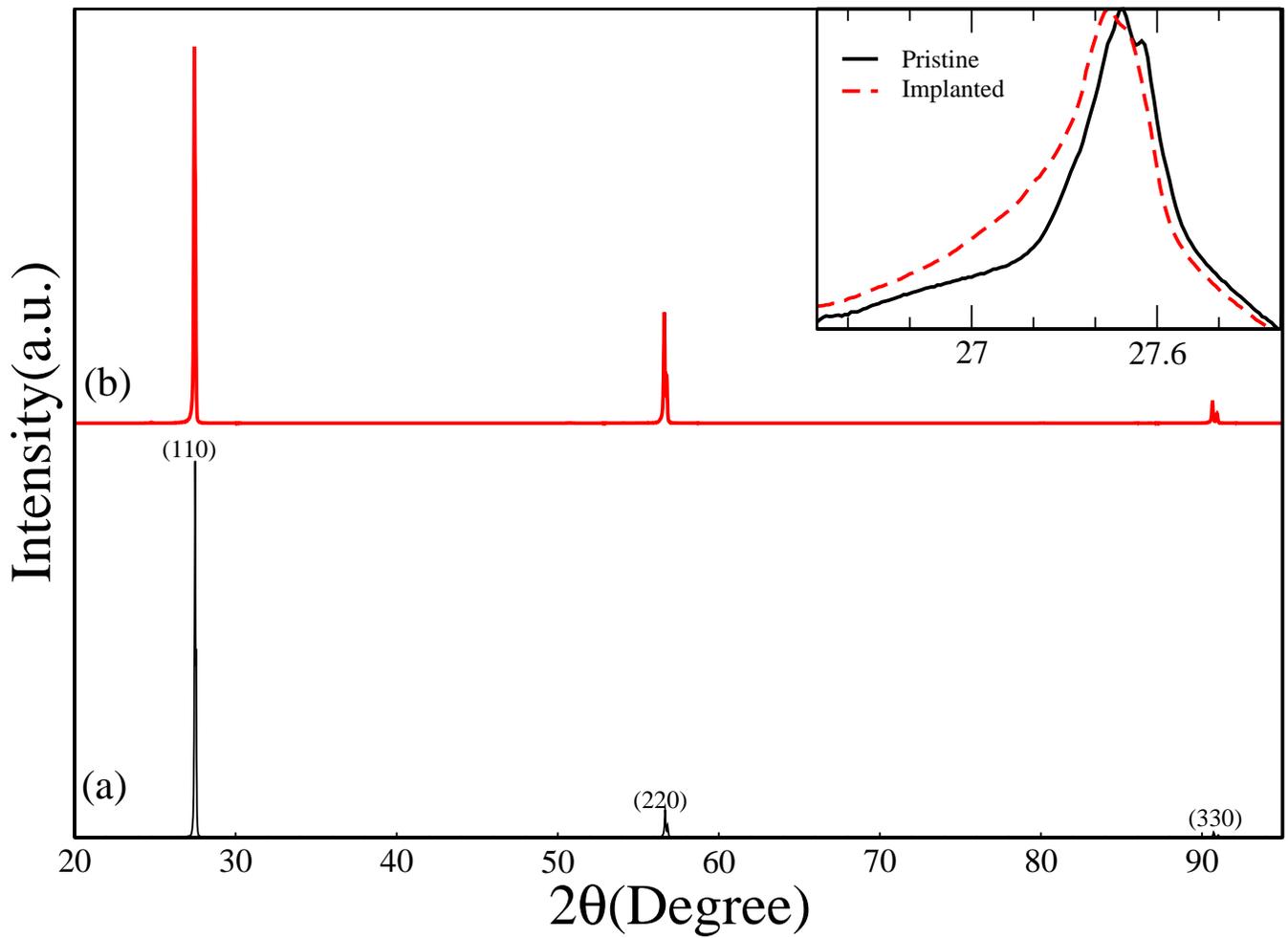}
\caption{X-Ray Diffraction (XRD) pattern of rutile TiO$_{2}$ for (a) pristine and after (b) cobalt implantation. Inset shows
the shift and peak broadening in TiO$_{2}$ (110) peak after cobalt implantation.}
\label{fig:xrd}
\end{figure}

\fig{fig:xrd} displays the XRD results for TiO$_{2}$ (110) samples both, prior to and after, implantation. 
The XRD diffraction pattern for pristine TiO$_{2}$ shows a very prominent peak at 27.5$^{\circ}$ which 
corresponds to (110) plane of rutile TiO$_{2}$ \cite{She1}. The other two lower intensity peaks at 56.7$^\circ$ and 
90.7$^\circ$ can be attributed to (220) and (330) planes. Though all these features are observed after
Co implantation in TiO$_{2}$ also, they are shifted towards lower diffraction angle (inset \fig{fig:xrd} displays
for (110) plane).
These shifts suggest an increase in the interplanar lattice spacing after implantation. This can be caused by the
substitution of some cobalt atoms in the TiO$_{2}$ lattice, altering the lattice parameter of the latter
due to the large ionic radii of cobalt atoms. This can also lead to the formation of compounds like 
Ti$_{1-x}$Co$_{x}$O$_{2}$ \cite{Lau}. Development of oxygen vacancies or Ti interstitials, during irradiation, 
can also cause modification in the lattice parameter as well as can generate lattice strain in the crystal \cite{Don1}.
Inset of \fig{fig:xrd} also displays a slight increase in the FWHM of the (110) feature for the implanted sample
as compared to the pristine. In addition, a prominent tailing effect is also observed in the lower angular region of 
this (110) feature (inset of \fig{fig:xrd}). These are related to the phenomenon of diffuse scattering indicating 
loss in crystallinity and enhancement of defects in the implanted sample, that disturb the long range 
ordering of crystals \cite{Mor}.

\begin{figure}
\includegraphics[width=0.5\textwidth]{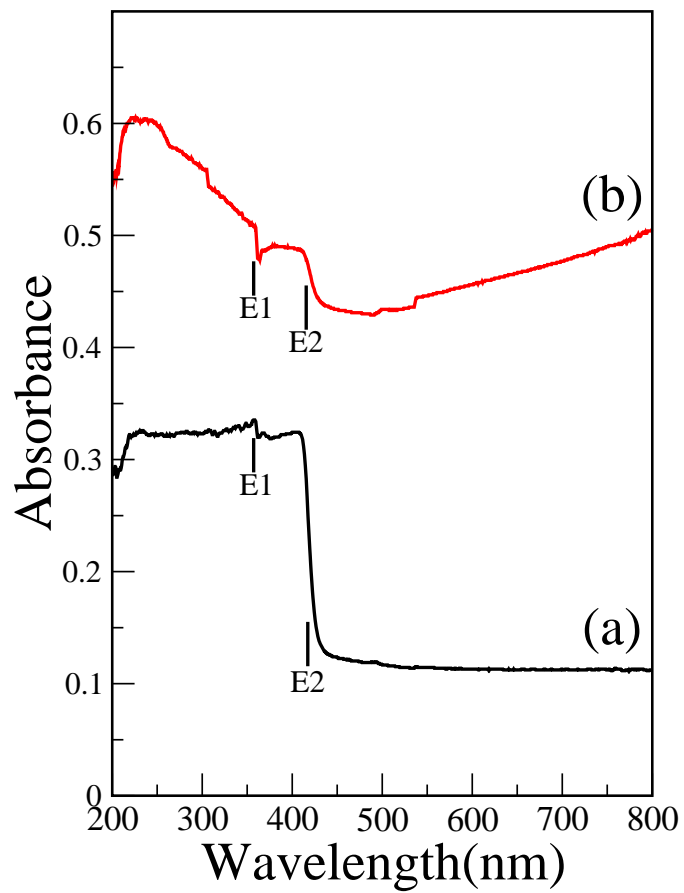}
\caption{UV-vis absorbance spectra of rutile TiO$_{2}$ for (a) pristine, and after (b) cobalt implantation.}
\label{fig:absorbance}
\end{figure}

The photoabsorption studies in the UV-vis regime were performed for the pristine as well as Co implanted rutile TiO$_{2}$ and
the results are displayed in \fig{fig:absorbance}. The pristine surface exhibits two band edges as shown in \fig{fig:absorbance}(a). E1, is the direct
band gap which arises due to optical transitions from O (2p) valance band to Ti (3d) conduction band while E2 is due to defects
present in the crystal \cite{Rum}. A significant enhancement in photoabsorption, about twice in UV regime and nearly five times
in visible regime, has been observed from the ion implanted TiO$_{2}$ surfaces (\fig{fig:absorbance}(b)). This enhancement of UV-vis photoabsorption after
Co implantation, is significantly higher than observed earlier for other dopants \cite{Che}. This enhancement
can be related to 
the incorporation of cobalt atoms in TiO$_{2}$ lattice which transfer their excess electrons to the conduction band of the
host lattice via metal to conduction band charge transfer process \cite{Klo}. 
In addition, formation of anisotropic nanostructures as well as creation of oxygen vacancies or Ti$^{3+}$ states on the ion implanted 
TiO$_{2}$ surfaces also  promote absorbance \cite{Ser}. The oxygen vacancies can act as trapping centers for electrons, with the
mean free path of electrons getting significantly reduced. This inhibits the electron-hole recombination process and hence promotes
the photoabsorbance \cite{Luo}. 

\begin{figure}
\includegraphics[width=0.5\textwidth]{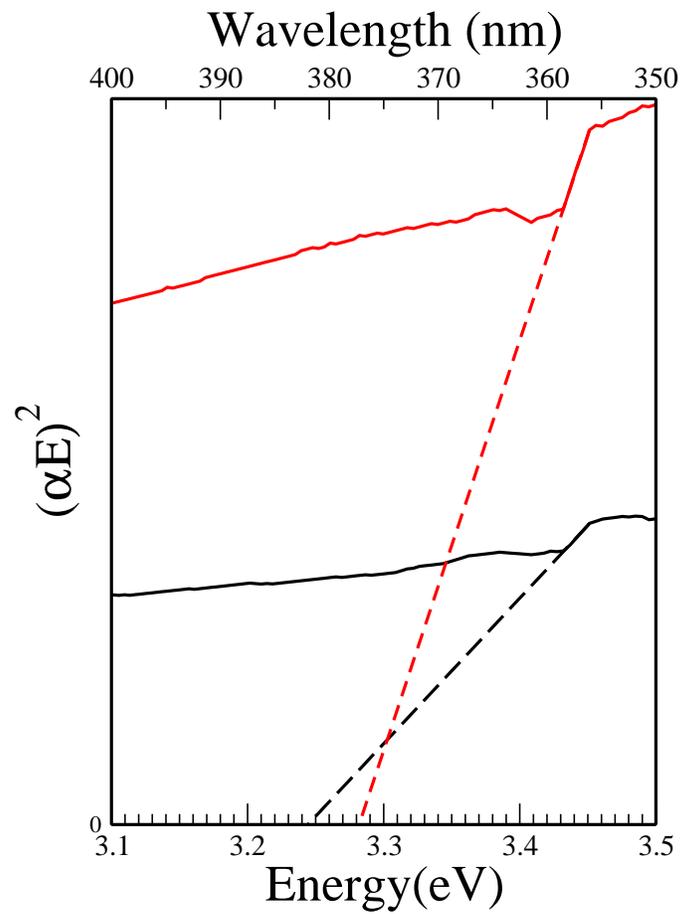}
\caption{Tauc plots of rutile TiO$_{2}$ for (a) pristine, and after (b) cobalt implantation. $\alpha$ is the
 absorption coefficient while E is the energy of the variable source.}
\label{fig:tauc}
\end{figure}

Photoabsorbance results of \fig{fig:absorbance} were utilized to obtain the Tauc plots shown in \fig{fig:tauc}. These have been utilized to study the band
gap variation for both, pristine and Co implanted TiO$_{2}$. The direct band gap of the pristine rutile TiO$_{2}$ is observed to be 
at 3.25~eV. This is slightly higher than the theoretically 
predicted band gap (3.06~eV) \cite{Pas}. After implantation this band gap increases very slightly to 3.28~eV. This redshift in direct 
band gap can be due to the quantum confinement effects, which according to the thermodynamic studies are
pronounced for nanostructures of sizes smaller than $\sim$30~nm \cite{Sat}. 

\section{Conclusion}
\label{conc}

In conclusion, we have investigated the photoabsorption properties of TiO$_{2}$ surfaces after implantation with cobalt ions.
Implantation leads to the fabrication of nanostructures as well as anisotropic ripple patterns. Development of oxygen vacancy
states, or Ti$^{3+}$ states,  is also observed on ion implanted surfaces. These surfaces show an enhanced photoabsorption, nearly 
twice and five times, in UV and visible regimes, respectively. Formation of nanostructured patterns and development of oxygen vacancies, 
upon Co implantation, lead to the enhanced photoabsorption observed here. The nanostructured
TiO$_{2}$ surfaces investigated here can have wide applications in photo- catalysis.

\section*{Acknowledgment}
We would like to acknowledge the help of Ramesh Chandra (IIT, Roorkee) for XRD measurements. Authors would
also like to acknowledge the help of Devrani Devi during Implantation at IUAC and Santosh Kumar Choudhury for XPS experiments.
\newpage
{\bf{ Figure caption }}\\
\begin {enumerate}
\item AFM images $(1\mu m\times 1\mu m)$ of rutile TiO$_{2}$ for (a) pristine, and after (b) cobalt implantation. 
Crystallographic directions of the surface are shown on pristine surface and remain same for the
implanted sample. Few nanostructures, on the implanted surface, are marked with solid circles.

\item (a) Diameter, and (b) Height distributions of nanostructures that develop on TiO$_{2}$ surface after cobalt implantation.  Solid 
lines represent Lorentzian fittings to the distributions.

\item XPS Core level spectra of Ti (2p) of rutile TiO$_{2}$ for (a) pristine surface, and after (b) cobalt implantation. Fittings of
components are shown.

\item X-Ray Diffraction (XRD) pattern of rutile TiO$_{2}$ for (a) pristine and after (b) cobalt implantation. Inset shows
the shift and peak broadening in TiO$_{2}$ (110) peak after cobalt implantation.

\item UV-vis absorbance spectra of rutile TiO$_{2}$ for (a) pristine, and after (b) cobalt implantation.

\item Tauc plots of rutile TiO$_{2}$ for (a) pristine, and after (b) cobalt implantation. $\alpha$ is the
 absorption coefficient while E is the energy of the variable source.

\end{enumerate}

\centerline{\bf References}
\centering

\newpage

\clearpage

\end{document}